\documentclass[prl,aps,superscriptaddress,twocolumn]{revtex4}
\usepackage{hyperref}
\usepackage{color} 
\usepackage{bm}
\usepackage{dcolumn}
\usepackage{times}
\usepackage{amssymb} 
\usepackage{amsmath} 
\usepackage{ragged2e}
\usepackage{graphicx}
\usepackage{epstopdf}
\usepackage{cancel}
\usepackage[english]{babel}
\usepackage{mathrsfs}
\usepackage{textcomp}
\usepackage[normalem]{ulem}
\usepackage{tikz}
\usepackage{pgfplots}
\usepgfplotslibrary{groupplots}

\AfterEndEnvironment{figure}{\noindent\ignorespaces}

\usepackage{amsthm}
\usepackage{amsfonts}
\usepackage{braket} 

\usepackage{cleveref}
\begin{document}

\title{Flux Magnetism in a Strongly Interacting Dipolar Lattice Supersolid under Tunable Gauge Fields}
\author{Michele Miotto}
\affiliation{Institut für Physik und Astronomie, Technische Universität Berlin, Hardenbergstraße
36, 10623 Berlin, Germany}
\author{Pietro Lombardi}
\affiliation{Consiglio Nazionale delle Ricerche - Istituto Nazionale di Ottica, sede secondaria di Sesto Fiorentino}
\affiliation{European Laboratory for Nonlinear Spectroscopy (LENS), Università di Firenze, Firenze, Italy}
\author{Giovanni Ferioli}
\affiliation{Department of Physics and Astronomy, University of Florence, 50019 Sesto Fiorentino, Italy}
\author{Joana Fraxanet}
\affiliation{ICFO - Institut de Ciencies Fotoniques, The Barcelona Institute of Science and Technology, Av. Carl Friedrich Gauss 3, 08860 Castelldefels (Barcelona), Spain}
\author{Maciej Lewenstein}
\affiliation{ICFO - Institut de Ciencies Fotoniques, The Barcelona Institute of Science and Technology, Av. Carl Friedrich Gauss 3, 08860 Castelldefels (Barcelona), Spain}
\affiliation{ICREA, Pg. Llu\'is Companys 23, 08010 Barcelona, Spain}
\author{Luca Tanzi}
\affiliation{Consiglio Nazionale delle Ricerche - Istituto Nazionale di Ottica, sede secondaria di Sesto Fiorentino}
\affiliation{European Laboratory for Nonlinear Spectroscopy (LENS), Università di Firenze, Firenze, Italy}
\author{Luca Barbiero}
\affiliation{Institute for Condensed Matter Physics and Complex Systems,
DISAT, Politecnico di Torino, I-10129 Torino, Italy}

\date{\today}

\begin{abstract}

Supersolidity and magnetism are fundamental phenomena characterizing strongly correlated matter. Here we unveil a mechanism that directly connects these two regimes and can be experimentally accessed in ultracold atomic systems. Specifically, we exploit the distinctive properties of magnetic lanthanide atoms trapped in a one-dimensional antimagic wavelength optical lattice. This platform enables a realistic implementation of a triangular Bose–Hubbard ladder featuring two key ingredients: strong long-range interactions and tunable gauge fields. Owing to these properties, our numerical analysis reveals a robust lattice supersolid regime with finite fluxes in each triangular plaquette. Remarkably, we show that the density modulation of the supersolid phase and a finite gauge field induce magnetic ordering of the fluxes, forming ferromagnetic and ferrimagnetic patterns. Our results thus reveal a quantum effect that bridges supersolidity and magnetism.

\end{abstract}

\maketitle

\paragraph{Introduction.}
The quantum mechanical effects underlying the formation of strongly interacting states of matter represent a fundamental topic in quantum physics~\cite{thouless2013,girvin2019}, yet key mechanisms remain unexplored. In this context, ultracold atomic systems~\cite{Lewenstein07,Lewenstein12,gross2017} have emerged as an innovative tool to deepen the understanding of a wide variety of quantum many-body phases~\cite{Goldman16,Arguello2019,Zhang02102018,Aidelsburger22}, among which magnetism~\cite{Buschow2003,Auerbach2012} is a primary example. Evidence for this is provided by the remarkable realizations of ferromagnetic~\cite{Parker2013,Cominotti2023}, antiferromagnetic~\cite{Mazurenko17,shao2024}, and ferrimagnetic~\cite{lebrat2024} states, where specific atomic degrees of freedom effectively reproduce electron spin orderings.\\
The fundamental role played by trapped atoms at ultralow temperatures~\cite{Goral2002,Recati2023} has been further highlighted by the observation of one of the most elusive states of matter: supersolidity~\cite{Andreev1969,Chester1970,Leggett1970}. 
While first instances of this regime have been achieved in different setups~\cite{Li2017,Leonard2017,chisholm2024}, the 
anisotropic long-range dipolar interaction promoted magnetic lanthanide atoms~\cite{Chomaz_2023} as the primary constituents of supersolid phases~\cite{tanzi2019,guo2019,chomaz2019}.\\ 
Great efforts have also been devoted to create strongly correlated phases driven by dipolar interactions in settings where magnetic atoms are trapped in optical lattices~\cite{dePaz2013nqm,Baier2016,Lepoutre2019ooe,Patscheider2020cde}. In this direction, important results have exploited a drastic reduction of the lattice spacing~\cite{Baier2016,Su2023,Du2024}. Nevertheless, accessing and probing these regimes remains highly challenging, and alternative strategies might be of crucial relevance.\\
\begin{figure}[h!]
    \centering
    \includegraphics[width=0.9\linewidth]{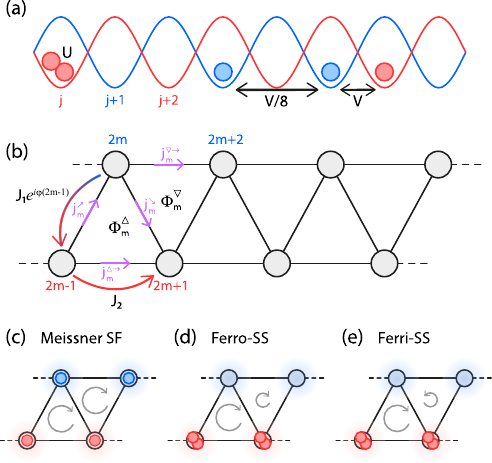}
    \caption{(a) Two-component gas of Dy trapped in a one-dimensional optical lattice at the antimagic wavelength and relative interactions, see first row of Eq.~\eqref{ham1}. We note that the $\lambda/2$ periodicity of the intensity profile corresponds to $\Delta j=2$. (b) Effective triangular ladder description of the model with correspondent hopping processes, see second and third rows of Eq.~\eqref{ham1}. Each triangular plaquette is characterized by effective current flux $\Phi_m^{\triangle}$ or $\Phi_m^{\bigtriangledown}$, see Eq.~\eqref{flux}, with currents indicated in violet and defined in Eq.~\eqref{currents}. (c)-(d)-(e) Local density and current flux patterns for the three observed phases in the case of explicitly broken time reversal symmetry.}
    \label{fig:exp}
\end{figure}
In this paper, we establish a connection between magnetism and supersolidity, which can be directly explored using an innovative experimental setup that significantly magnifies the long-range dipolar repulsion.
In particular, we first design a configuration where magnetic dysprosium atoms are confined in a one-dimensional antimagic wavelength optical lattice, which enables two Zeeman states to occupy spatially shifted sublattices separated by $\lambda/4 \sim 100$~nm. Raman coupling~\cite{Lin2011,celi2014,Mancini2015,Stuhl2015} between spin states generates tunable complex tunneling processes across the sublattices, effectively synthesizing a triangular extended Bose-Hubbard ladder in a gauge field~\cite{Struck2013,Anisimovas2016,Halati2023,Baldelli2024,chanda2024,Li2020}. 
Crucially, at this subwavelength scale the dipolar interaction between nearest-neighbor sites becomes comparable to the on-site one in a broad range of lattice depths.\\
Our numerical analysis reveals that, because of these intriguing features, interesting many-body phases, including a lattice supersolid~\cite{Goral2002,Batrouni2013,Li2013,Mishra2015,Kraus2022}, characterize the ground state of this system. Strikingly, this supersolid regime hosts staggered flux patterns with magnetic order—either ferromagnetic or ferrimagnetic—arising from the interplay between finite gauge fields and the dipolar repulsion induced density modulation. 
Finally, we provide a concrete procedure to prepare and probe the intriguing states of matter that our results unveil.

\paragraph{Magnetic atoms in an antimagic wavelength optical lattice.}
\begin{figure}
    \centering
    \includegraphics[width=.8\linewidth]{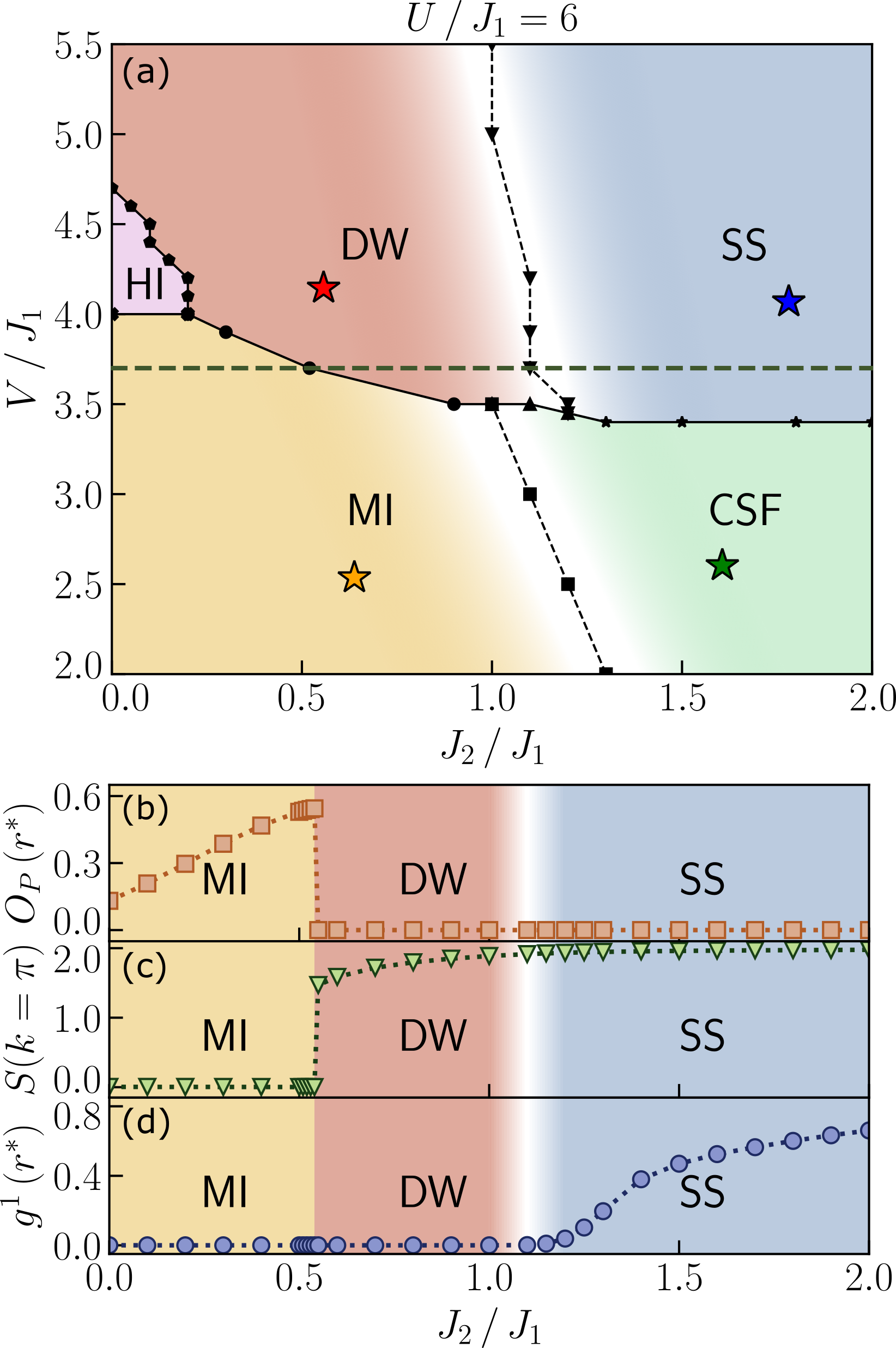}
    \caption{(a) Phase diagram of the model in Eq. \eqref{ham1} at $\varphi=\pi$ and $U/J_1=6$, as a function of $V/J_1$ and $J_2/J_1$ for total density $\bar{n}=1$. Here, we observe five different phases: Mott insulator (MI), Haldane insulator (HI), density wave (DW), chiral superfluid (CSF) and lattice supersolid (SS). The stars refer to concrete experimental parameters, see \cite{SM} for details, giving rise to the indicated values of $J_2/J_1$, $U/J_1$, $V/J_1$ and $\varphi$. Dashed lines and color gradients signal transitions of the Kosterlitz-Thouless (KT) type, while solid lines indicate first-order transitions. The horizontal dashed line at $V/J_1=3.7$ represents the cut along which the order parameters are shown in the panels below. (b) Parity operator~\eqref{eq:parity} as a function of $J_2/J_1$, computed at $r^*=200$. (c) Structure factor~\eqref{dw} for $k=\pi$ as a function of $J_2/J_1$. (d) Expectation value of $g^1$~\eqref{sf} as a function of $J_2/J_1$, computed at $r^*=200$. In the DMRG simulations, we used maximum bond dimension $\chi = 800$.}
    \label{fig:phaseU}
\end{figure}
Our scheme builds on magnetic lanthanide atoms, such as dysprosium and erbium, confined in a one-dimensional optical lattice of wavelength $\lambda$. These atoms exhibit a large tensorial polarizability in the ground state due to their unfilled $4f$ shell, giving rise to strongly state-dependent optical potentials~\cite{Li2016,Becher2018,Chalopin2018, Bloch2023,Du2024}. 
By exploiting the ability to tune the scalar-to-tensorial polarizability ratio near an electronic resonance, an antimagic potential~\cite{Anisimovas2016,Baldelli2024} for two successive Zeeman states can be achieved. In this regime, the selected states experience optical potentials of equal depth and opposite sign~\cite{SM}, thus producing two spin-selective sublattices displaced by $\lambda/4$. Importantly, this enables interparticle spacings as small as $100$~nm—well below standard optical lattice scales—without resorting to ultraviolet light~\cite{Sohmen2023,Fraxanet2022}. 
We focus on bosonic dysprosium atoms in the Zeeman sublevels $\ket{a} = \ket{J=8, m_J=-8}$ and $\ket{b} = \ket{J=8, m_J=-7}$, which are particularly suitable as they exhibit suppressed dipolar relaxation and nearly equal contact interactions $U_{aa}\approx U_{bb}=U$ when the magnetic field is tuned near $2.75$~G~\cite{Lecomte2025}. The antimagic condition for $\ket{a}$ and $\ket{b}$ is satisfied whenever the scalar and tensorial polarizabilities reach a specific ratio, which naturally occurs near electronic resonances~\cite{SM}. In our scheme, this condition is fulfilled in the $529.87$–$530.25$~nm range. Crucially, in this configuration the effective nearest-neighbor dipolar repulsion reaches $V/h \approx 130$~Hz, thus comparable to the on-site interaction $U$ across a wide range of lattice depths~\cite{SM}. Note that similar dipole-dipole interaction strengths are obtained with polar molecules in lattices with $\sim 0.5~\mu$m spacing~\cite{Christakis2023,Carroll2025}. 
Coherent tunneling between the two spin states is induced by a two-photon Raman coupling~\cite{Lin2011,Burdick2016}, which drives spin-flip hopping between adjacent sites of the two sublattices. This process generates a complex nearest-neighbor tunneling term $J_1 e^{i\varphi}$, where the amplitude is set by the Raman intensity and spatial overlap between the initial and the final states, while the gauge field $\varphi$ is determined by the beam geometry~\cite{Lin2011,celi2014,Mancini2015,Stuhl2015}. In contrast, next-nearest-neighbor tunneling $J_2$ arises from kinetic processes within each sublattice and is controlled by the lattice depth. The resulting system is described by the extended Bose-Hubbard Hamiltonian
\begin{align}\label{ham1}
H&=\sum_{j=1}^{L}\Big[V\sum_{k>j}^{L}\frac{n_j n_{k}}{(k-j)^3}+\frac{U}{2}n_{j}(n_j-1)\Big] \notag\\
&-\sum_{m=1}^{(L-2)/2} \Big[J_{2} \left(a_{2m-1}^\dagger a_{2m+1} + a_{2m}^\dagger a_{2m+2} + \text{H.c.}\right)\\
&+J_{1}\left(e^{\imath\varphi(2m-1)}a_{2m-1}^\dagger a_{2m}+e^{-\imath\varphi(2m)}a_{2m}^\dagger a_{2m+1}+\text{H.c.}\right)\Big]\notag .
\end{align}

Notably, although our approach relies on a purely one-dimensional trapping potential, Eq.~\eqref{ham1} effectively describes a strongly interacting triangular ladder in a gauge, see Fig.~\ref{fig:exp}. More in detail, here $a^\dagger$ and $a$ are the usual bosonic operators. For ease of notation, the index $m$ runs over the effective triangular plaquettes and is shared by two consecutive plaquettes, while $j$ and $k$ run over all $L$ lattice sites. The specific structure of the nearest-neighbor hopping term leads to triangular plaquettes being pierced by a gauge field of $\pm \varphi$. In the following Sections, we employ the density-matrix renormalization group (DMRG) method~\cite{White1992, Schollwoeck2011, tenpy2024} to investigate the possible quantum phases emerging from the distinctive properties of $H$. 
\paragraph{Geometrically frustrated regime ($\varphi=\pi$).}
We begin our analysis by fixing $\varphi=\pi$ and the particle density $\bar{n}=\frac{N_a+N_b}{L}=1$. Notably, this choice generates a sign staggered real-valued $J_1$, which, for finite $J_2$, can induce effective geometric frustration ~\cite{Eckardt_2010,Cabedo2020,Roy2022,Halati2023,Barbiero2023,Baldelli2024,Halati2025,Burba2025}.
Our results in Fig.~\ref{fig:phaseU}(a) show that for small $J_2/J_1$ the frustration is not effective and Eq.~\eqref{ham1} reproduces the phase diagram of the widely investigated frustration-free 1D extended Bose-Hubbard model~\cite{dallatorre2006,Deng2011,Rossini_2012,Batrouni2013,Ejima2015,Kottmann2020,Fraxanet2022}. In this regime, dominant onsite repulsions $U$ stabilize a Mott insulator (MI), captured by the parity order parameter~\cite{Berg2008,Endres2011}
\begin{equation}
    O_P \left( r \right) = e^{i \pi \sum_{j<r} \delta n_j} \: \: 
    \label{eq:parity}
\end{equation}
with $\delta n_j=(\bar{n}-n_j)$, see Fig.~\ref{fig:phaseU}(b). 
By increasing $V$, the MI is replaced by the topological Haldane insulator (HI), while for even stronger dipolar repulsions, a density wave (DW) insulator emerges. In the latter, the translational symmetry breaks down and alternation between sites with high and low occupation occurs. As is intuitive, this density modulation is captured by a finite peak at $k=\pi$ in the structure factor   
\begin{equation}
S(k)=\frac{1}{L}\sum_{r}e^{-\imath k r}\langle n_jn_{j+r}\rangle.
\label{dw}
\end{equation}
Importantly, increasing geometric frustration, i.e., taking larger $J_2/J_1$ ratios, profoundly alters this picture. Specifically, for low $V$ the MI undergoes a transition of the Kosterlitz-Thouless (KT) type and is replaced by a chiral superfluid (CSF)~\cite{SM} with spontaneously broken time-reversal symmetry~\cite{Dhar2012,Roy2022,Halati2023,Barbiero2023,Baldelli2024,Burba2025}.
In Fig.~\ref{fig:phaseU}(a), see the horizontal dashed line, we further find that for larger dipolar repulsions, geometric frustration can induce a first-order phase transition from a Mott insulating to a density wave regime. Evidence of this is reported in Figs.~\ref{fig:phaseU}(b) and (c), where we show that at a critical value $J_2/J_1$ the parity operator $O_P \left( r \right)$ vanishes and the finite value $S(k=\pi)$ signals the breaking of the translational symmetry. 
As shown in Fig.~\ref{fig:phaseU}(d), important information can also be extracted from the behavior of the single particle Green function 
\begin{equation}
g^1(|k-j|)=\langle b_k^\dagger b_{j}\rangle
\label{sf}
\end{equation}
for finite large distances $|k-j|$. In both MI and DW, $g^1(|k-j|)\approx0$ already for relatively short distances, indicating an exponential decay distinctive of insulating phases. In contrast, by increasing $J_2/J_1$ the single particle Green function decays algebraically and its value for large $|k-j|$ becomes significant. According to Luttinger theory~\cite{Giamarchi2004}, this behavior signals the onset of a one-dimensional gapless superfluid (SF) phase. Crucially, in this regime $S(\pi)\neq0$, indicating that superfluidity coexists with the breaking of the translational symmetry. As a consequence, these results define a lattice supersolid (SS) phase. The DW-SS transition is also of the KT type, analogous to the MI-SF transition observed in the conventional 1D Bose-Hubbard model \cite{Kuhner2000}.
It is important to emphasize that, in the regimes of relatively large $U$, this SS phase cannot take place if either $J_2\approx0$ or $V\approx0$. This highlights that the combination of geometric frustration and large dipolar repulsions is responsible for the appearance of this $\bar{n}=1$ lattice SS phase. 
\paragraph{Magnetic ordering of fluxes.}
\begin{figure}
    \centering
    \includegraphics[width=.8\linewidth]{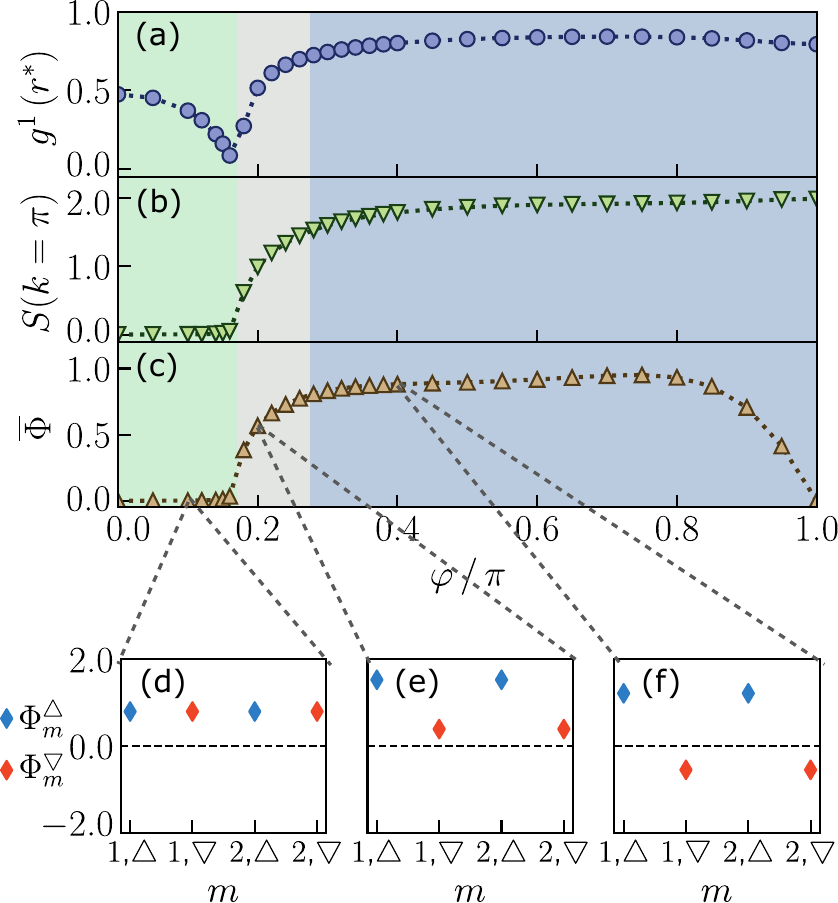}
    \caption{Behavior of (a) the single particle Green function $g^1$ at $r^*=60$; (b) the structure factor $S(\pi)$; (c) the overall staggered flux average $\bar{\Phi}$, as a function of $\varphi/\pi$ and fixed $U/J_1=6$, $V/J_1=4.5$ and $J_2/J_1=1.7$. Green regions correspond to the Meissner superfluid, while grey and violet areas indicate ferro- and ferrimagnetic supersolid phases, respectively. Panels (d)-(e)-(f) show the effective current fluxes per plaquette $\Phi_{m}^{\triangle}$ and $\Phi_{m}^{\bigtriangledown}$ for $\varphi/\pi=\left\{0.1,0.2,0.4\right\}$, respectively. The plaquettes are taken in the middle of the chain of length $L=101$. In the DMRG simulations, we used maximum bond dimension $\chi = 1000$. We have checked that the values of the currents entering the equations for $\Phi^{\triangle}_m$ and $\Phi^{\bigtriangledown}_m$ are robust against changes of the length of the chain.}
    \label{fig:ss}
\end{figure}
Except for a few known examples~\cite{Han2018,Yevtushenko2018}, SS states typically preserve time-reversal invariance. Here, the tunability of the gauge field enables us to systematically investigate the properties of supersolidity under explicitly broken time-reversal symmetry. Important insight on this latter scenario can be also unveiled through the behavior of the generated current flux in odd and even effective triangular plaquettes
\begin{align}
    &\Phi_{m}^{\triangle} = j^{\nearrow}_m + j^{\searrow}_{m} - j^{\triangle\rightarrow}_m,\nonumber\\
    &\Phi_{m}^{\bigtriangledown} = -j^{\nearrow}_{m+1} - j^{\searrow}_{m} + j^{\bigtriangledown\rightarrow}_m
    \label{flux}
\end{align}
respectively, where 
\begin{align}
&j^{\nearrow}_{m}=-\imath J_1\left(e^{\imath\varphi(2m-1)}a_{2m-1}^\dagger a_{2m}-\text{H.c.}\right)\notag ,\\
&j^{\searrow}_{m}=-\imath J_1 \left( e^{-\imath\varphi(2m)}a_{2m}^\dagger a_{2m+1}-\text{H.c.}\right)\notag ,\\
&j^{\triangle\rightarrow}_m=-\imath J_2\left(a^{\dagger}_{2m-1} a_{2m+1}-\text{H.c.}\right)\notag ,\\
&j^{\bigtriangledown\rightarrow}_m=-\imath J_2\left(a^{\dagger}_{2m} a_{2m+2}-\text{H.c.}\right)
\label{currents}
\end{align}
are the currents along the diagonal and horizontal links, see Fig. \ref{fig:exp}(b), and the overall staggered flux average~\cite{SM1}
\begin{align}
\bar{\Phi}=\sum_{m=1}^{(L-3)/2}(\Phi_{m}^{\triangle}-\Phi_{m}^{\bigtriangledown})/(L-3).
\label{avg_flux}
\end{align}
In our analysis, we fix the Hamiltonian parameters $U$, $V$, $J_1$, and $J_2$ able to stabilize the SS regime illustrated in Fig.~\ref{fig:phaseU}(a), and explore the effects produced by variations of the gauge field in the range $0 < \varphi < \pi$.
In Fig.~\ref{fig:ss}(a), we report that, for a weak $\varphi$, superfluidity — signaled by $g^1(r^*)\neq0$ — persists, while the vanishing $S(\pi)$ in Fig.~\ref{fig:ss}(b) indicates that translational invariance is preserved~\cite{SM2}. Therefore, we can conclude that supersolidity is unstable for small gauge fields and a homogeneous superfluid emerges as the favorable ground state.
We now exploit the similarity between the periodic structures that $\Phi_{m}^{\triangle}$/$\Phi_{m}^{\bigtriangledown}$ can develop 
and the magnetic ordering of spins in the odd/even sites of a 1D chain with effective staggered magnetization $\bar{\Phi}$. Based on this, the results in Figs.~\ref{fig:ss}(c) and (d) find this SF phase, usually called Meissner SF~\cite{Piraud2015,Halati2023,Impertro2025}, to be characterized by an effective ferromagnetic flux structure as $\Phi_{m}^{\triangle}=\Phi_{m}^{\bigtriangledown}>0$ and $\bar{\Phi}=0$, see also \cite{SM}. By increasing $\varphi$, different flux magnetic orderings emerge. Specifically, Figs.~\ref{fig:ss}(a) and (b) report that for $\varphi\geq0.18\pi$ both $g^1(r^*)$ and $S(\pi)$ are finite, thus a first-order transition related to the spontaneous breaking of the translation symmetry occurs and a SS state takes place. Fig.~\ref{fig:ss}(c) further explains that this SS phase features an average staggered flux $\bar{\Phi}>0$ that signals the asymmetric alternation between $\Phi_{m}^{\triangle}$ and $\Phi_{m}^{\bigtriangledown}$. Importantly, this latter condition is induced by the presence of a finite gauge field and by the SS density structure. Indeed, the density imbalance between the legs causes the emergence of finite interleg current, which is equal along all diagonal links $j_m^{\nearrow}=j_m^{\searrow}$. Furthermore, the high/low occupation of odd/even sites implies $\left|j^{\triangle\rightarrow}_m\right|>\left|j^{\bigtriangledown\rightarrow}_m\right|$~\cite{SM}. These facts, together with $j^{\triangle\rightarrow}_m<0$, lead to $\Phi_{m}^{\triangle}>\Phi_{m}^{\bigtriangledown}>0$~\cite{SM3} in this range of parameters, see Fig.~\ref{fig:ss}(e). As a consequence, we can identify such a phase as a supersolid characterized by a ferromagnetic flux structure (Ferro-SS). For larger $\varphi$, our numerics show that the density modulation of the SS phase still enables the appearance of a staggered flux pattern. Importantly, we observe a crossover to the conditions $\Phi_{m}^{\triangle}>0>\Phi_{m}^{\bigtriangledown}$ and $|\Phi_{m}^{\triangle}|>|\Phi_{m}^{\bigtriangledown}|$ being fulfilled, as shown in Fig.~\ref{fig:ss}(f). In analogy with spin chains, this structure thus describes a ferrimagnetic flux ordering embedded in a SS phase (Ferri-SS).\\
These results thus unveil that intriguing states of matter — the Ferro-SS and Ferri-SS phases  — where supersolidity and magnetism coexist can characterize strongly correlated regimes. Importantly, these unconventional features emerge thanks to the specific density structure of a lattice SS phase combined with large enough gauge fields.
\paragraph{State preparation and detection scheme.}
Here, we show how these many-body phases can be experimentally prepared and probed with high accuracy. Specifically, the ground state of the Hamiltonian Eq.~\eqref{ham1} can be prepared by directly loading a Dy Bose-Einstein condensate in the $\ket{8,-8}$ state into the antimagic optical lattice. With an appropriate tuning of interatomic interaction and axial confinement, the system can be initialized in a low-energy, doubly occupied, Mott insulator phase. Adiabatically turning on the Raman coupling induces a finite occupation of the $\ket{8,-7}$ state~\cite{SM}, thereby enabling access to the unit-filling regime $\bar{n}=1$. Moreover, as accurately shown in~\cite{SM} and reported in Fig.~\ref{fig:phaseU}(a), by exploiting the possibility to independently tune all the Hamiltonian parameters $J_1$, $\varphi$, $J_2$, $U$, and $V$, the discussed regimes turn out to be experimentally accessible. Taking advantage also of the synthetic nature of the ladder geometry, 
such states of matter can be detected in the experiment using standard spin-resolved time-of-flight (TOF) absorption imaging, without the need for a quantum gas microscope and single-site resolution. Indeed, TOF images provide information about 
the momentum distribution of the atomic sample in the 
lattice. On the one hand, this measurement gives access to the single-particle Green function $g^1(r)$ \cite{Greiner2002}, thus distinguishing phases with superfluid properties (SF and SS) from insulators (MI and DW). On the other hand, by performing state-sensitive 
imaging, e.g. by applying a magnetic field gradient during the free expansion, one can differentiate between phases characterized by mainly one spin component (SS and DW) and phases characterized by two equally populated spin components (SF and MI). 
To detect the different magnetic orderings appearing in the SS regime,  
an investigation of the currents in both the real and the synthetic dimensions is required. A simple measurement in state resolved band-mapping procedure is enough to determine the presence of collective currents in the real one~\cite{Mancini2015}. Concerning the synthetic direction,
following~\cite{Impertro2024}, we propose to project the ladder system  
onto a 1D lattice of isolated double wells 
in the synthetic dimension, each well corresponding to a spin state, and use the complex hopping $J_1$ induced by the Raman transition to map the current operator between the two wells 
onto the double-well population imbalance, see~\cite{SM}. 
\paragraph{Conclusions and perspectives.}
Our results unveil a mechanism linking supersolidity and magnetism. Thanks to the capability of achieving strong dipolar repulsions in the presence of tunable gauge fields, our derived experimental setup offers a direct path to probe such a scenario as well as to explore strongly correlated regimes dominated by long-range interactions. As we prove, these configurations can be realistically achieved by exploiting the properties of Raman coupled magnetic lanthanide atoms trapped in an optical lattice at the antimagic wavelength. Motivated by the versatility offered by this innovative setup, our numerical analysis reveal that the interplay of tunable gauge fields and large local and long-range interactions makes possible the appearance of a lattice supersolid phase where fluxes order into ferromagnetic or ferrimagnetic structures.\\     
From an experimental perspective, our results provide a powerful and alternative strategy to magnify the strength of the dipolar interaction in lattice systems without resorting to UV wavelengths~\cite{Sohmen2023,Fraxanet2022} or bilayer lattice schemes~\cite{Du2024}. 
The further tunability of gauge fields, makes it possible to envision our setup as an important future tool for exploring strongly interacting phases of matter in effective magnetic fields, whose quantum simulation represents a challenging and timely research subject~\cite{Zhou2023,Impertro2025}. In addition, because of the controlled spin state occupation, this architecture can be further employed to engineer Hamiltonians in constrained spin space~\cite{Claude2024,douglas2024,Lecomte2025}. We finally remark that, although we concentrated on dysprosium, our scheme directly applies to erbium.\\
From a theory perspective, a natural question concerns the possible existence of similar linking mechanisms between different strongly correlated states of matter. In addition, 
 an important future direction will be to explore whether similar flux orderings emerge in 2D lattices, where supersolidity is also predicted~\cite{Capogrosso2010}. Importantly, magnetically ordered fluxes in 2D can also suggest the presence of topologically protected phases~\cite{Giuliani_Vignale_2005c}. As a consequence, our results might represent a first preliminary step to reveal a possible connection between supersolidity and topology.\\
Our results provide a new avenue to create and understand many-body regimes dominated by long-range interactions and reveal an intriguing interplay between supersolidity and magnetism.\\

Calculations were performed using the TeNPy library (version 1.0.0)~\cite{tenpy2024}.\\
\paragraph{Acknowledgments.}
We thank N. Baldelli, C. Cabrera, S. Dhar, A. Eckardt, F. Ferlaino, M. Landini, M. Mark, G. Modugno, L. Santos and G. Valtolina for discussions. M. M., L. T. and L. B. acknowledge funding from the Italian MUR (PRIN DiQut Grant No.~2022523NA7). M. M. acknowledges funding from the Deutsche Forschungsgemeinschaft (DFG, German Research Foundation) via the Research Unit FOR 5688 (Project No. 521530974). P. L., G. F. and L. T. acknowledge funding from the European Union (European Research Council, SUPERSOLIDS, Grant No. 101055319). M. L. acknowledges support from: European Research Council AdG NOQIA; MCIN/AEI (PGC2018-0910.13039/501100011033, CEX2019000910-S/10.13039/501100011033, Plan National FIDEUA PID2019-106901GB-I00, Plan National STAMEENA PID2022-139099NB, I00, project funded by MCIN/AEI/10.13039/501100011033 and by the “European Union NextGenerationEU/PRTR”; (PRTRC17.I1), FPI); QUANTERA DYNAMITE PCI2022132919, QuantERA II Programme co-funded by European Union’s Horizon 2020 program under Grant Agreement No 101017733; Ministry for Digital Transformation and of Civil Service of the Spanish Government through the QUANTUM ENIA project call - Quantum Spain project, and by the European Union through the Recovery, Transformation and Resilience PlanNextGenerationEU within the framework of the Digital Spain 2026 Agenda; Fundaci\'o Cellex; Fundaci\'o Mir-Puig; Generalitat de Catalunya, European Social Fund FEDER and CERCA program; Barcelona Supercomputing Center MareNostrum (FI-2023-3-0024); funded by the European Union; (HORIZON-CL4-2022-QUANTUM-02-SGA PASQuanS2.1, 101113690, EU Horizon 2020 FETOPEN OPTOlogic, Grant No 899794, QU-ATTO, 101168628), EU Horizon Europe Program (This project has received funding from the European Union’s Horizon Europe research and innovation program under grant agreement No 101080086 NeQSTGrant Agreement 101080086 — NeQST); ICFO Internal “QuantumGaudi” project.

Views and opinions expressed are however those of the author(s) only and do not necessarily reflect those of the European Union, European Commission, European Climate, Infrastructure and Environment Executive Agency (CINEA), or any other granting authority. Neither the European Union nor any granting authority can be held responsible for them\\

\paragraph{Data availability.} The data that support the findings of this article are openly
available~\cite{ZenodoData}.

\bibliographystyle{apsrev4-2}
\bibliography{biblio.bib}

\newpage
\section{Supplemental Material}

\subsection{Antimagic optical potential}
In general, the dynamical polarizability of an atom can be decomposed into three different contributions known as scalar, vectorial and tensorial polarizability \cite{LepersErPol,Li2016,Becher2018}. Since the realization of the many-body phases described in the main text requires repulsive long-range interactions,  we focus here on a specific configuration where the magnetic field is perpendicular to the propagation axis of the trapping light, thus maximizing the dipolar repulsion between the atoms. In this configuration, the vectorial  polarizability is zero, and the potential created by light on an atom is then proportional to the sum of two terms:
\begin{eqnarray}
\begin{split}
     V(r, \omega)=-\frac{1}{2c\epsilon_0} I(r) \big[\alpha_s(\omega) +
     g(m_J)f(\theta_p)\alpha_t(\omega) \big]
\end{split}
\end{eqnarray}
where $g(m_J)=\frac{3 m_J^2-J(J+1)}{2J(J-1)}$ and $f(\theta_p)=\frac{3\cos^2\theta_p -1}{2}$, $\epsilon_0$ is the vacuum permittivity, $c$ is the speed of light, $\omega$ is the frequency of the trapping light, and $\theta_p$ represents the angle between the polarization axis of the laser field and the quantization axis, which is generally set by the magnetic field.

\begin{figure}
    \centering
    \includegraphics[width=\linewidth]{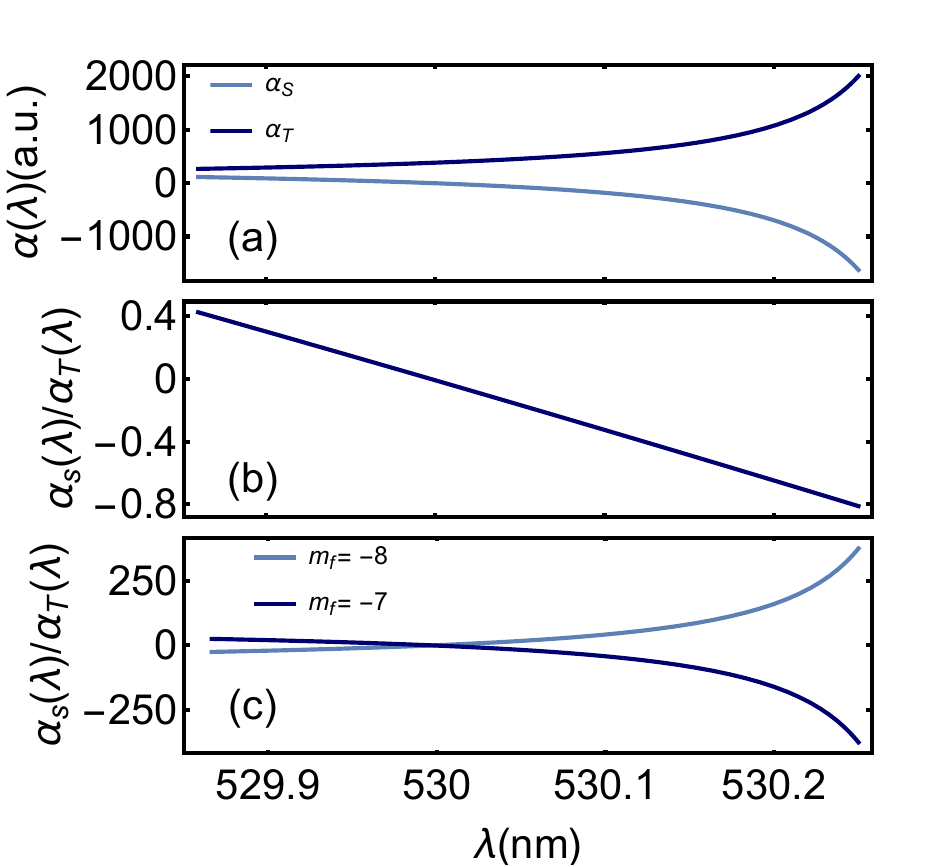}
    \caption{Polarizability evaluated using the transitions theoretically predicted in \cite{DzubaDyPol, Li2016}, in proximity of 530~nm. (a) Scalar and tensorial contributions for the ground state $\ket{8,-8}$. Their ratio is instead in (b). In (c) the total polarizability of the two internal states $\ket{8,-8}$ and $\ket{8,-7}$ is reported.}
    \label{fig:SupMat3}
\end{figure}

Unlike alkali atoms, for lanthanides $\alpha_t$ is relevant even in the ground state due to their open $f$ shell structure. Thus, the light-induced trapping potential depends on both the atomic Zeeman sublevel $m_J$ and the polarization of the electromagnetic field.
Our key idea is to exploit the strong tensor contribution to the polarizability to create a controllable state-dependent potential for dysprosium atoms in the Zeeman sublevels $\ket{a}$ and $\ket{b}$, defined as $\ket{a}=\ket{J=8,m_J=-8}$ and $\ket{b}=\ket{J=8,m_J=-7}$. In particular, we find that an antimagic potential can be generated for states $\ket{a}$ and $\ket{b}$ if 
\begin{eqnarray}
\frac{\alpha_S}{\alpha_T}(\omega)=-\frac{g(-8)+g(-7)}{2}f(\theta_p)=-0.8125 f(\theta_p),
\end{eqnarray}
where $f(\theta_p)$ ranges between $1$ and $-0.5$ if $\theta_p$ varies between $0$ and $\pi/2$. 
Therefore, the antimagic condition can be achieved for a given range of laser frequencies $\omega$ whenever $\alpha_s(\omega)$ crosses zero, that is for laser wavelengths close to an electronic transition. Here, we have chosen to work near the electronic transition located at $\lambda = 530.307$~nm, a wavelength short enough to have a small lattice spacing and narrow enough to have a sufficiently low scattering rate, but other choices are possible. In particular, we identify the antimagic condition in the range $529.87-530.25$~nm, depending on $\theta_p$, as shown in Fig.~\ref{fig:SupMat3}. For the specific antimagic configuration at $\theta_p = 0$ and $\lambda =530.25$~nm, we find $\alpha=378$~au. Therefore, $s= V_L/E_{R} \simeq 15$ can be achieved with about $40$~mW of lattice power considering a beam waist of $50$~$\mu$m, conditions that are realistically achievable with commercial lasers. We have also evaluated that the photon scattering rate remains low at this intensity, $\Gamma_{sc}<1$~Hz.

These calculations have been carried out considering the transitions reported and theoretically predicted in \cite{Li2016, DzubaDyPol}. However, precise spectroscopic data for the transition located at $\lambda = 530.307$~nm are still missing \cite{igorPol}, and the exact value of the antimagic configuration requires experimental validation.

\subsection{Raman coupling}
In this section, we briefly discuss the main experimental parameters describing the Raman coupling between spin states $\ket{a}=\ket{8,-8}$ and $\ket{b}=\ket{8,-7}$, which represents the tunneling in the synthetic dimension $J_1$. A suitable choice for the Raman coupling is the $741$~nm transition between the ground state $4f^{10}6s^2(^5I_8)$ and the excited state $4f^95d6s^2(^5K^o_9)$, which has a linewidth of $1.8$~kHz. Assuming a detuning from resonance of $\Delta = 1$~GHz and a laser power of about $25$~$\mu$W distributed over a laser beam of $100$~$\mu$m waist, it is possible to estimate an on-site effective Rabi frequency for the Raman transition $\Omega_R=\Omega_1\Omega_2/\Delta\simeq 400$~Hz. Here, $\Omega_{1,2}$ are the Rabi frequencies of the transition between $\ket{a}\rightarrow\ket{9,-8}$ ($\pi$ polarized) and $\ket{9,-8}\rightarrow\ket{b}$ ($\sigma^-$ polarized), respectively. This configuration maximizes the Clebsch-Gordan coefficient and hence the Raman frequency.

On top of the Zeeman shift related to the bias magnetic field, the detuning between the two Raman beams must be set to precisely compensate for the differential AC Stark shift induced by the optical potential and by the Raman beams themselves.
In our configuration, the first term corresponds to the lattice depth $s=V_{Lat}/E_R$ and is of the order of $65$~kHz for $s=15$. The second term can be neglected as it is estimated to be of the order of $10$~Hz and hence smaller than the Raman Rabi frequency.

\begin{figure}!
    \centering
    \includegraphics[width=\linewidth]{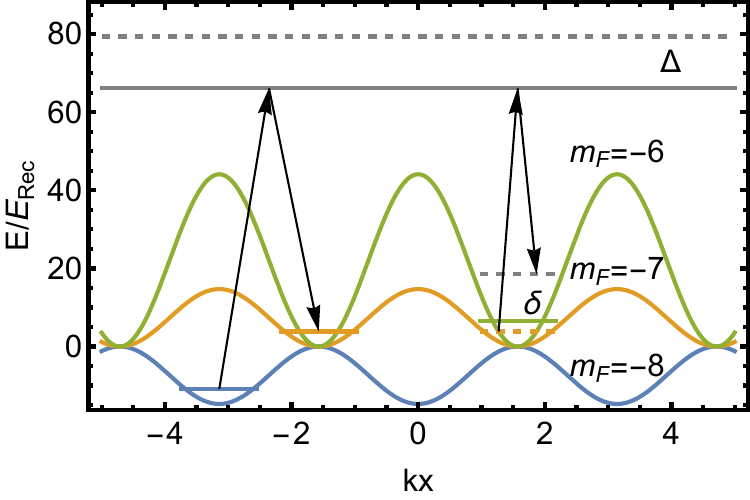}
    \caption{Energy levels of the three internal states $\ket{8,-8}$, $\ket{8,-7}$ and $\ket{8,-6}$, considering a specific value of the lattice realizing the supersolid phase described in the main text ($s\simeq 13$). The black arrows represent the Raman coupling between $\ket{8,-8}$ and $\ket{8,-7}$. Due to the different polarizability of $\ket{8,-6}$, the Raman coupling to this state is detuned by $\delta$.  }
    \label{fig:SupMat1}
\end{figure}

We note that, as shown in Fig. \ref{fig:SupMat1}, the Stark shift that has to be compensated for in the $\ket{8,-8}\rightarrow\ket{8,-7}$ Raman transition is very different from that in the $\ket{8,-7}\rightarrow\ket{8,-6}$ Raman transition. Remarkably, this fact automatically prevents the occupation of all Zeeman states with $m_J\ge -6$, isolating the $\ket{a}$ and $\ket{b}$ states in an effective two-level system.
More precisely, since the polarizability of the $\ket{8,-6}$ state is about three times that of the $\ket{8,-7}$ state, at $s=15$ the energy difference between the ground states of the $\ket{8,-6}$ and the $\ket{8,-7}$ optical lattices is $E_{rec}/h(\sqrt{3s}-\sqrt{s})\simeq 12$~kHz $\ll65$~kHz.
The Raman beams are therefore not resonant to this process. The detuning required to couple the ground state of the $\ket{8,-7}$ optical potential and the first excited band of the $\ket{8,-6}$ optical potential is instead about $70$~kHz, but we expect the Raman coupling to be suppressed given the opposite parity of the two wavefunctions.
Moreover, the $5$~kHz difference between the ground state transition $\ket{8,-8} \rightarrow \ket{8,-7}$ and the intraband one $\ket{8,-7} \rightarrow \ket{8,-6}$ is about an order of magnitude larger than the Rabi frequency considered in this work, which would lead to an effectively small population transfer even in the case of equal parity.

\subsection{Calculations of the Hamiltonian parameters}
In this section, we provide additional details related to the experimental parameters required for the exploration of the phase diagram reported in Fig.~\ref{fig:phaseU}. We assume a 1D system with a radial harmonic confinement of $\omega_{\perp}\simeq 2\pi 5$~kHz, resulting in a harmonic-oscillator length $l_{\perp}=(\hbar/m\omega_{\perp})^{1/2}\simeq 110$~nm. The main parameters of the extended Bose-Hubbard model can be evaluated considering the following integral:
\begin{equation}
    U_{i,j,k,l}= \int dx dx'w_i^*(x)w_i^*(x')V_{1D}(|x-x'|)w_i(x)w_l(x')
\label{Eq:U}
\end{equation}
where the functions $w_i(x)$ represent the localized Wannier function. Assuming a strong radial confinement and the tight binding regime, they can be approximated to Gaussian possessing a width equal to $l_{\perp}$ in the radial directions and to $2E_R\sqrt{s}/\hbar$ along the lattice direction, where $E_R = 2\pi^2\hbar^2/(m\lambda^2)$ is the recoil energy associated to an optical lattice with wavelength $\lambda$. The expression in Eq.\, \ref{Eq:U} can then be evaluated considering the quasi one directional interacting potential \cite{Fraxanet2022}:
\begin{eqnarray}
\begin{split}
      \frac{V_{1D}(|x-x'|)}{E_R}=\big{(}g_{1D}-\frac{2\hbar^2 r^*}{3 m l_{\perp}^2}\big{)}\delta(|x-x'|) +\\ +\frac{2\hbar^2 r^*}{m l_{\perp}^3}
    \Bigg[\sqrt{\frac{\pi}{8}}e^{|x-x'|^2/l_{\perp}^2}\Big{(} \frac{|x-x'|^2}{l_{\perp}^2}+1\Big{)}\times\\\times\text{Erfc}\Big{(} \frac{|x-x'|}{l_{\perp}\sqrt{2}}\Big{)} -\frac{|x-x'|}{2l_{\perp}} \Bigg]
\end{split}
\end{eqnarray}
where $r^*= m \mu_0 \mu^2/4\pi\hbar^2$ indicates the dipolar length, while $g_{1D}=2\hbar^2 a_{3D}/ml_{\perp}^2$ accounts for the contact interaction. The on-site interaction is evaluated as
$U=U_{i,i,i,i}$, while the interaction between nearest-neighbor sites is $V=U_{i,i+1,i,i+1}$. For the specific antimagic configuration considered in this work $w_{i+1}(x)=w_i(x+\lambda/4)$. Once fixed the transverse confinement and the wavelength adopted for the optical lattice, all the other quantities can be evaluated as a function of $s$: the tunneling is in fact given by $J_2/E_R = \frac{4}{\sqrt{\pi}}s^{3/4}e^{-2\sqrt{s}}$, while the tunneling in the synthetic dimension depends on the Rabi frequency associated to the Raman coupling and by the spatial overlapping between wavefunctions located in nearest-neighbor sites, $J_1= \Omega_{Raman}\int dx w_{i}(x)w_{i}(x+\lambda/4)$. 

\begin{figure}
    \centering
    \includegraphics[width=\linewidth]{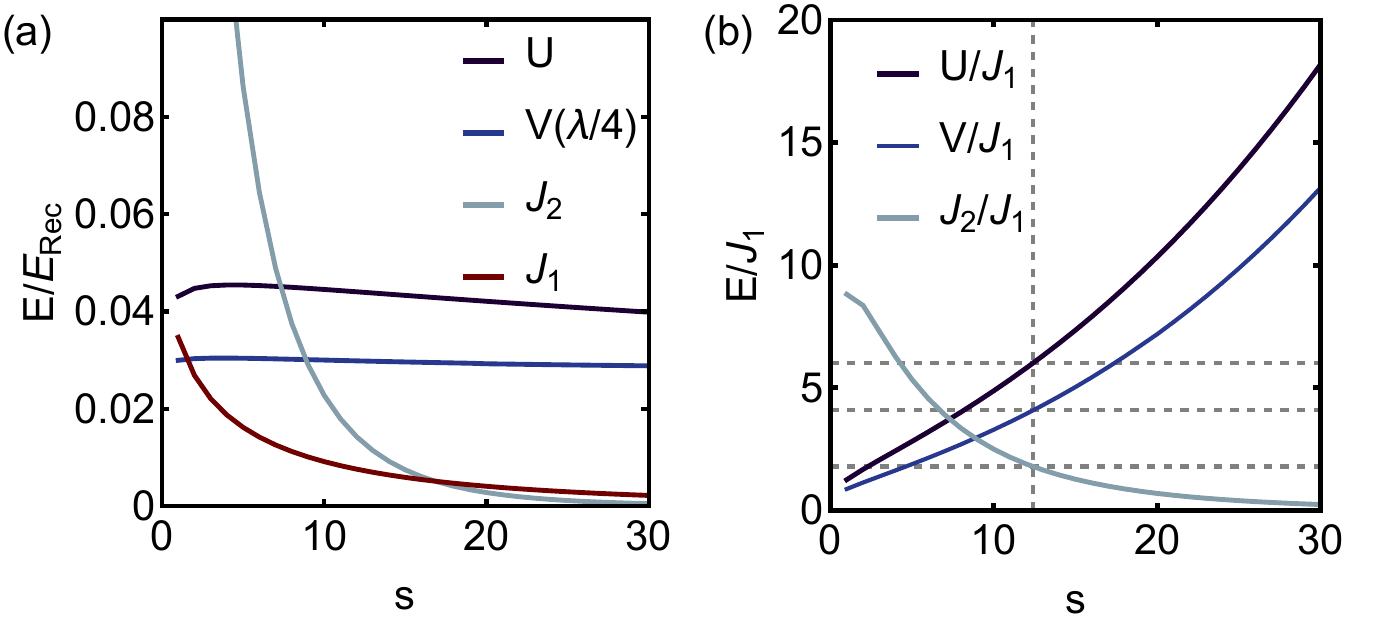}
    \caption{Scaling of the main parameters used in the extended Bose-Hubbard model as a function of the lattice depth $s$. For this specific case, evaluated using an external confinement of $\omega_\perp = 2\pi 5$~kHz and $\Omega_R = 280$~Hz, $U=6 J_1$ for $s\simeq13$. At this depth, $V/J_1\simeq 4.1$ while $J_2/J_1\simeq 1.8$, resulting in the SS phase.  }
    \label{fig:SupMat2}
\end{figure}

The scaling of $U$, $V$, $J_1$ and $J_2$ with $s$ is reported in Fig.\,\ref{fig:SupMat2}(a). Interestingly, $U$ does not depend strongly on the lattice depth $s$. This behavior is not present in systems limited to contact interaction and is originated by the anisotropic and long-range nature of dipolar interaction. \color{black}

\begin{figure}
    \centering
    \includegraphics[width=\linewidth]{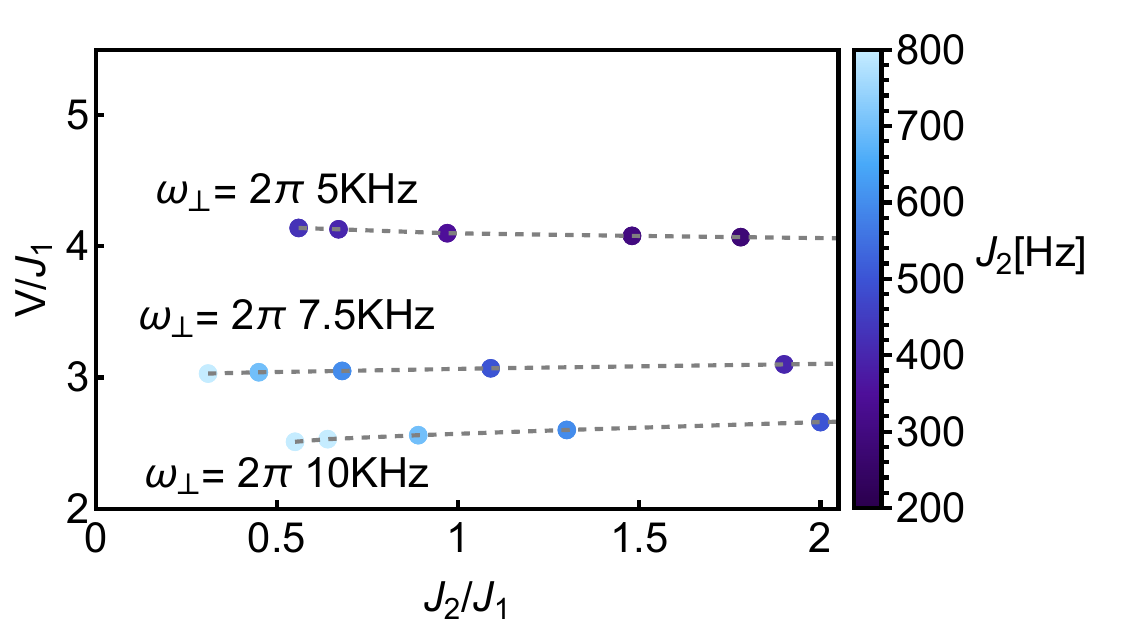}
    \caption{ 
    Different values of $V/J_1$ and $J_2/J_1$ fulfilling the condition $U/J_1= 6$ evaluated for different values of $\omega_\perp$ and $\Omega_R$. The latter is indicated by the color scale of each point, reported in the color bar.}
    \label{fig:SupMat2b}
\end{figure}

To estimate the experimental parameters required for exploring the phase diagram in Fig.\,\ref{fig:phaseU}, we proceed as follows. We set the radial confinement and we evaluate $U$, $V$, $J_1$ and $J_2$ as a function of $s$, assuming a variable Rabi frequency of the Raman transition $\Omega_R$. Then we calculate the specific value of $s$ required to satisfy the condition $U= 6 J_1$ and estimate for that value $J_1/J_2$ and $V/J_2$, as shown in Fig.\,\ref{fig:SupMat2}(b). In order to scan the phase diagram, we can change either $\Omega_R$ or the external radial confinement. We report, in Fig.\,\ref{fig:SupMat2b}, the experimental conditions required to explore the phase diagram described in the main text, evaluated with the protocol just explained. \color{black}
\subsection{Limits to the coherent evolution time given by dipolar relaxation}
As recalled in the main text, unless the spatial confinement overcomes the Zeeman splitting, coherent spin evolution in dipolar quantum gases is severely limited by dipolar relaxations \cite{pasquiou2011spontaneous}. Indeed, the intrinsic coupling between internal and external angular momentum translates magnetic depolarization into atom losses if the corresponding gain in kinetic energy is greater than the confining potential, and this condition generally occurs even for magnetic fields as low as few mG. According to the relaxation rate reported in literature for bosonic dysprosium \cite{burdick2015fermionic, Lecomte2025}, a superfluid phase in a lattice with average density close to unity collapses within $1$~ms. In the proposed scenario, however, the effective relaxation rate can be strongly reduced thanks to the weak transversal confinement and the staggered geometry provided by the antimagic lattice, with an estimated overall lifetime beyond $10$~ms. Taking advantage of the magnetic field "sweet spot" described in \cite{Lecomte2025}, the coherence time can be further extended by two orders of magnitude, hence largely exceeding the $100$~ms time scale required by the adiabatic loadings designed in the current proposal. 

\subsection{State preparation}
The initial state for the exploration of the new phases described in the paper is a single spin component doubly occupied Mott insulator. The realization of this state can be achieved by different strategies. For example, a set of independent Bose-Einstein condensates in an almost 1D system can be created using optical tweezers or an accordion lattice, both of which can provide the required transverse confinement of about $5$~kHz while maintaining negligible tunneling in the transverse plane. The formation of a doubly occupied Mott insulator in the case of a cloud of about 100 atoms and axial confinement of $50$~Hz is achievable for realistic lattice depth and bias magnetic field.

We note that this condition can also be promoted by applying a repulsive potential capable of shaping the effective harmonic confinement along the axial direction at will. Similar techniques have already been successfully demonstrated, for example in \cite{Mazurenko17, gall2021competing}.

After such preparation, the magnetic field and the lattice depth have to be set to the values of interest ($s\simeq 15$, for the lattice depth). These ramps can be performed during the adiabatic switch-on of the Raman coupling, required to populate the ground state of the effective triangular lattice,  which is expected to last for around $100$~ms.

\subsection{State detection}
\paragraph{k-vector distribution measurement via TOF imaging.}

Sufficiently long free-falling expansion before absorption imaging allows a direct mapping of the k-vector distribution at the instant of release into the spatial distribution of the imaged single-particle wavefunction. The overlap among the contributions from the different atoms, on one hand enables the detection even in case of limited number of atoms ($< 10^4$), and on the other hand gives access to the inter-site coherence of the system manifesting as an interference pattern.
Combined with the implementation of state-sensitive detection, such imaging technique is sufficient for the identification of the different phases introduced in the paper, with the exclusion of the local current properties.

\paragraph{Currents in the real and synthetic dimensions.}
In order to assess the non-trivial plaquette flux ordering of the SS phase in the proposed system, an investigation of the currents in both the real and the synthetic dimensions is required. Concerning the former, a simple measurement in state resolved band-mapping configuration is enough to determine the presence of collective currents. The evaluation of currents in the synthetic ribbons is more challenging. In \cite{Impertro2024, Impertro2025}, Impertro and coworkers present and implement an original protocol that allows the probe of local kinetic operators (both kinetic energy and current) via standard local occupation measurements. 
The implementation of such protocol in our system is possible provided some specifically developed technical steps. In particular, the triangular ladder geometry can be transformed into an array of double wells thanks to the following sequence: 
1) turn off of the Raman coupling and rise of the 1D lattice depth to around $s=25$; 
2) introduction of a state-independent optical linear gradient of the order of few kHz/$\mu$m (attainable for instance exploiting the transverse intensity profile of a far-detuned  $100$~$\mu$m-diameter laser beam (e.g. $1064$~nm wavelength with a power of around $1$~W).
In these conditions, the hopping in real space $J_2$ is limited to few Hz, while the Raman coupling $J_1$ can be restored for an adjusted laser beam intensities and detuning, as the spatial overlap between the two states is still not extinguished. However, because of the optical gradient, Raman-driven hopping is limited to only one of the next-neighbor sites at a time, thus realizing the double-well array. 
One tricky point of the protocol consists in determining the correct phase of the $\pi/2$ Raman coupling pulse to probe the current rather than the kinetic energy. In fact, while the new optimal parameters (frequency difference and pulse area) can be determined by a characterization of the system in the new configuration, the phase of the coherence between the two components in the double well is determined by the temporal integral of the difference in energy between the two states during the whole process. Even though this phase is not fully predictable, a high stability of the experimental setup can guarantee the reproducibility of the phase trajectory, thus opening to a tomographic probing of the kinetic energy-current plane, as discussed in \cite{Impertro2024}. 

\subsection{Robust supersolidity for fixed $J_2/J_1$}

\begin{figure}
    \centering
    \includegraphics[width=.8\linewidth]{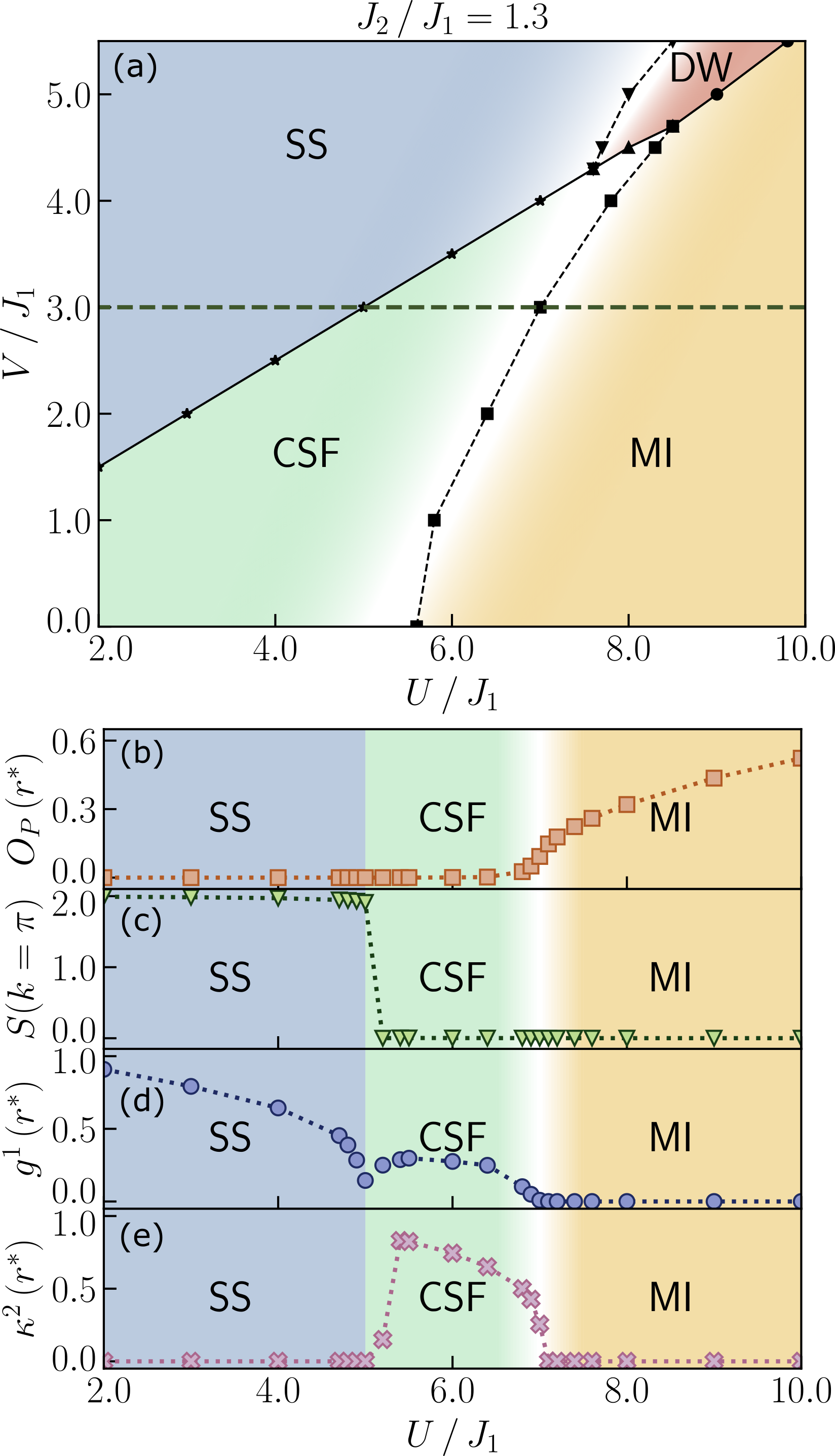}
    \caption{(a) Phase diagram of the model in Eq.~\eqref{ham1} at $\varphi=\pi$ and $J_2/J_1=1.3$, as a function of $V/J_1$ and $U/J_1$ for total density $\bar{n}=1$. Dashed lines and color gradients signal transitions of the KT type, while solid lines indicate first-order transitions. The horizontal dashed line at $V/J_1=3.0$ represents the cut along which the order parameters are shown in the panels below. (b) Parity operator~\eqref{eq:parity} as a function of $U/J_1$, computed at $r^* = 200$. (c) Structure factor~\eqref{dw} for fixed $k=\pi$ as a function of $U/J_1$. (d) Expectation value of $g^1$~\eqref{sf} as a function of $U/J_1$, computed at $r^* = 200$. (e) Chiral correlator~\eqref{chiral_correlator} as a function of $U/J_1$, computed at $r^* = 200$. In the DMRG simulations, we used maximum bond dimension $\chi = 800$.}
    \label{fig:phaset2}
\end{figure}

In addition to the analysis of the geometrically frustrated regime for fixed onsite interaction $U/J_1$, which is discussed in the main text, we also performed DMRG simulations in the case of fixed $J_2/J_1=1.3$. The results shown in Fig.~\ref{fig:phaset2} highlight that the same many-body phases also exist in the ground state of this setting, except for the HI, whose topological order is destroyed by the high levels of geometric frustration. Since the other phases have been already characterized in the main text, we focus here on the chiral superfluid (CSF). As shown in Fig.~\ref{fig:phaset2}(d), the latter exhibits a finite value of $g^1(r)$ for large distances like a conventional superfluid, but it is fully captured by the long-range order of the chiral correlation function
\begin{equation}
    \kappa^2 \left( \left| k-j \right| \right) = \langle \kappa_k \kappa_j \rangle,
	\label{chiral_correlator}
\end{equation}
where we have introduced the vector chiral order parameter
\begin{equation}
    \kappa_j = - \frac{\imath}{2} \left( b_j^{\dagger} b_{j+1} - b_{j+1}^{\dagger} b_j \right).
    \label{chiral_operator}
\end{equation}
The long range order of the chiral correlator~\eqref{chiral_correlator} in the CSF region in Fig.~\ref{fig:phaset2}(e) signals that such phase is characterized by alternated finite currents between nearest-neighbor sites, thus spontaneously breaking the time reversal symmetry of the model.
\vfill
\subsection{Currents in magnetically ordered phases}

\begin{figure}
    \centering
    \includegraphics[width=0.8\linewidth]{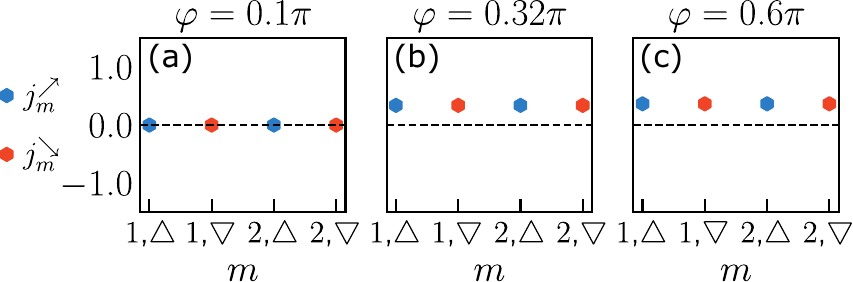}
\caption{(a)-(b)-(c) Values of current between nearest-neighbor sites for $\varphi/\pi=\left\{0.1,0.32,0.6\right\}$. The other parameters of Eq.~\eqref{ham1} are the same as in Fig.~\ref{fig:ss}, i.e. $U/J_1=6$, $V/J_1=4.5$ and $J_2/J_1=1.7$. Currents as defined in Eq. \eqref{currents}. For the Meissner SF in panel (a), all currents are vanishing.}
    \label{fig:ss_diag}
\end{figure}

\begin{figure}
    \centering
    \includegraphics[width=0.8\linewidth]{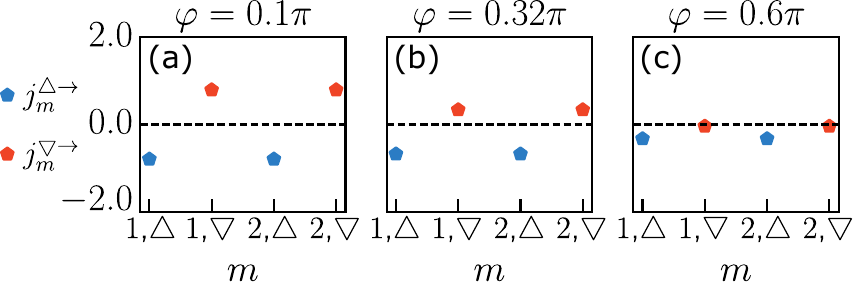}
    \caption{(a)-(b)-(c) Values of current between next-nearest-neighbor sites for $\varphi/\pi=\left\{0.1,0.32,0.6\right\}$. The other parameters of Eq.~\eqref{ham1} are the same as in Fig.~\ref{fig:ss}, i.e. $U/J_1=6$, $V/J_1=4.5$ and $J_2/J_1=1.7$. Currents as defined in Eq. \eqref{currents}. In panel (c), the red markers are at negative finite values.}
    \label{fig:ss_hor}
\end{figure}

We report here additional details on the current patterns in the case of explicitly broken time reversal symmetry. As discussed in the main text, when fixing $U$, $V$ and $J_2$ to produce the SS in the frustrated regime, for small $\varphi$ we find a Meissner SF. As shown in Figs.~\ref{fig:ss_diag}(a) and \ref{fig:ss_hor}(a), the latter is characterized by $j_m^{\nearrow} = j_m^{\searrow} = 0$ and $j_m^{\triangle \rightarrow} = - j_m^{\bigtriangledown \rightarrow}$. The currents in the two SS phases are instead more interesting to analyze. They both exhibit $j_m^{\nearrow} = j_m^{\searrow} > 0$ along the diagonal links, as reported in Figs.~\ref{fig:ss_diag}(b) and (c), while a richer behavior occurs for currents between next-nearest-neighbor sites. Indeed, unlike any phase observed in \cite{Halati2023}, in the entire Ferro-SS and in part of the Ferri-SS region for $\varphi \leq 0.5\pi$, our results in Figs.~\ref{fig:ss_hor}(b) and (c) yield $j^{\triangle\rightarrow}_m<0<j^{\bigtriangledown\rightarrow}_m$. On the contrary, in the Ferri-SS for $\varphi > 0.5\pi$, we find $j^{\triangle\rightarrow}_m<j^{\bigtriangledown\rightarrow}_m<0$: here, the two horizontal currents move in the same direction. In both cases, we observe $\left|j^{\triangle\rightarrow}_m\right| > \left|j^{\bigtriangledown\rightarrow}_m\right|$, since we chose the configuration that favors occupation in the lower leg of the triangular ladder.

\end{document}